\documentclass[twocolumn]{aastex63}

\shorttitle{Formation of MSP PSR J1012+5307 with an ELM WD}
\shortauthors{N. Wei et al.}

\usepackage{amsmath}
\usepackage{graphicx}
\usepackage{float}
\usepackage{subfigure}
\begin{document}
	

\title{Formation of PSR J1012+5307 with an extremely low-mass white dwarf: testing magnetic braking models}


\author[0009-0006-3261-6418]{Na Wei}
 \affil{School of Science, Qingdao University of Technology, Qingdao 266525, People's Republic of China; chenwc@pku.edu.cn}
\author[0000-0002-9739-8929]{Kun Xu}
 \affil{School of Science, Qingdao University of Technology, Qingdao 266525, People's Republic of China; chenwc@pku.edu.cn}
\author[0000-0002-3007-8197]{Zhi-Fu Gao}
  \affil{Xinjiang Astronomical Observatory, Chinese Academy of Sciences, Urumqi, Xinjiang 830011, People's Republic of China}
\author[0000-0002-2479-1295]{Long Jiang}
  \affil{School of Science, Qingdao University of Technology, Qingdao 266525, People's Republic of China; chenwc@pku.edu.cn}
   \affil{School of Physics and Electrical Information, Shangqiu Normal University, Shangqiu 476000, People's Republic of China}
\author[0000-0002-0785-5349]{Wen-Cong Chen}
  \affil{School of Science, Qingdao University of Technology, Qingdao 266525, People's Republic of China; chenwc@pku.edu.cn}
  \affil{School of Physics and Electrical Information, Shangqiu Normal University, Shangqiu 476000, People's Republic of China}



\begin{abstract}
PSR J1012+5307 is a millisecond pulsar with an extremely low-mass (ELM) white dwarf (WD) companion in an orbit of 14.5 hours. Magnetic braking (MB) plays an important role in influencing the orbital evolution of binary systems with a low-mass ($<1-2~M_{\odot}$) donor star. At present, there exist several different MB descriptions. In this paper, we investigate the formation of PSR J1012+5307 as a probe to test the plausible MB model. Employing a detailed stellar evolution model by the MESA code, we find that the Convection And Rotation Boosted MB and the 'Intermediate' MB models can reproduce the WD mass, WD radius, WD surface gravity, neutron-star mass, and orbital period observed in PSR J1012+5307. However, our simulated WD has higher effective temperature than the observation. Other three MB mechanisms including the standard MB model are too weak to account for the observed orbital period in a Hubble time. A long cooling timescale caused by H-shell flashes of the WD may alleviate the discrepancy between the simulated effective temperature and the observed value.
\end{abstract}


\section{Introduction}
Extremely low-mass (ELM) white dwarfs (WDs) are low-mass helium (He) WDs with a mass of $\sim0.2~M_{\rm \odot}$ and a surface gravity in the range of $5 < \log g < 7$ \citep{Brow13}. The majority of ELM WDs are discovered in binary millisecond pulsars (MSPs) as the companion stars \citep{van05}. MSPs are recycled neutron stars (NSs) with a short spin period and a weak surface magnetic field \citep{Back82}, which were spun up to millisecond period by accreting matter and angular momentum from their companion stars, i.e. recycling process \citep{Alpa82,Radh82}. An ELM WD is the product of binary evolution in which the envelope of a low-mass donor star is stripped by the accreting NS leaving behind a light He WD \citep{taur96a,taur96b,istr14a}. The formation of ELM WDs is closely tied to some important but indistinct physical processes such as mass exchange and angular momentum loss. Therefore, it is important to study the formation and evolution of ELM WDs in the stellar and binary evolution field.

In the recycling model \citep{Alpa82,Radh82,bhat91}, the progenitors of binary MSPs are low-mass X-ray binaries (LMXBs) consisting of a NS and a low-mass donor star with a mass of $\sim 1-2~\rm M_{\odot}$. With a nuclear evolution of long timescale of $\sim 10^{9}~\rm yr$, the donor star fills its Roche lobe, and transfers the surface H-rich material onto the NS. The evolutionary endpoint is a binary MSP consisting of a MSP and a low-mass He WD after the mass accretion of $\sim 10^{9}~\rm yr$ \citep{liu11,shao12}. The donor star during the LMXB stage should experience a red giant stage due to the expansion of the envelope. There exists an ideal relation between the degenerate core mass and the radius of the giant for those low-mass stars ($\la2.3~\rm M_{\odot}$) \citep{refs71,webb83}. During the mass transfer, the radius of the giant equals the radius of its Roche lobe, which relates to the orbital separation \citep{eggl83}. After the exhaustion of the giant envelope, its He core gradually evolves into a WD by a contraction and cooling stage. As a consequence, a good correlation between the orbital periods and the masses of the WDs should exist for wide binary MSPs with orbital periods $P_{\rm orb}\ga 1~\rm days$ \citep{joss87,savo87,rapp95,taur99}. \cite{jia14} showed that this relation can be extended to narrow binary MSPs with an ELM WD ($0.14-0.17~\rm M_{\odot}$) and a short orbital period (several hours).

According to the correlation between the orbital period and the WD mass, an ELM WD corresponds to a tight orbital separation. PSR J1012+5307 was identified as a binary MSP (spin period of 5.26 ms) with a low-mass WD companion in a $P_{\rm orb} = 14.5~\rm hours$ long \citep{nica95} near-circular ($e<8.4\times10^{-7}$) orbit with a moderate inclination angle \citep{drie98,lang01,laza09}. Subsequently, the observations of the optical spectroscopy indicate that its companion star is an ELM WD \citep{lori95,kerk96,call98}. Based on the VLBI observations spanning $\sim2.5$ years for this source, its distance was measured to be $d=0.83^{+0.06}_{-0.02}~\rm kpc$ \citep{ding20}. The cross-correlation of the individual spectra with a synthetic WD template implies a radial velocity semi-amplitude of $K_{\rm 2}= 218.9\pm22~\rm {km~s^{-1}}$ and a systemic velocity of $\gamma=-21.3\pm1.6~{\rm km~s^{-1}}$ for PSR J1012+5307 \citep{mata20}. An accurate mass ratio of $q = 10.44 \pm 0.11$ was got by referring to the ephemerides derived from the precise radio timing. The spectral classification provided the best estimation of the effective temperature and surface gravity of the ELM WD to be $T_{\rm eff}=8362^{+25}_{-23}~\rm K$ and $\log g=6.26\pm0.04$, respectively. A comparison with photometric observations of the all-sky survey inferred the WD radius to be $R_{\rm WD}=0.047^{+0.003}_{-0.002}~R_{\rm \odot}$ \citep{mata20}. The detailed stellar evolutionary models for ELM WDs proposed a conservative WD mass to be $M_{\rm WD}=0.165\pm0.015~M_{\rm \odot}$. Employing the derived mass ratio and the inclination angle ($i=50\pm2~{\rm deg}$) of the source, it yields a NS mass as $M_{\rm NS}=1.72\pm0.16~M_{\rm \odot}$ \citep{mata20}.  So far, the formation of this MSP with an ELM WD has not yet been fully understood.

The formation of ELM WDs is related to the mass-transfer process, especially the mechanism of angular momentum loss. For a close binary with an orbital period less than 2 hours, the gravitational radiation dominates the mass transfer and orbital evolution. However, magnetic braking (MB) plays an important role in driving the mass transfer of systems with an orbital period longer than 3 hours. Therefore, the MB model is successful in accounting for the period gap of $2-3$ hours in the cataclysmic variables observations \citep{rapp83,kni11}. However, the simulated mass-transfer rates employing the standard MB prescription are at least an order of magnitude lower than the observations in LMXBs \citep{pods02}. Subsequently, several MB prescriptions were proposed (see also section 2) to interpret some observed properties in the stellar and binary evolution field. As a MSP with an ELM WD, PSR J1012+5307 is an ideal probe for testing the MB mechanism and the binary evolution theory.

This work aims to test the different MB models by investigating the formation of PSR J1012+5307. This paper is organized as follows. We describe the different MB models in section 2. Based on the different MB models, we perform a detailed stellar evolution model for the formation of PSR J1012+5307 to find the best MB model that can match its observations in Section 3. Finally, we give a brief discussion and summary in Sections 4 and 5.

\section{Different MB models}
In a binary system, the low-mass star would spin down due to the coupling between the stellar
winds and the magnetic field \citep{verb81}. However, the tidal interaction between the two components would
continuously accelerate the star to co-rotate with the orbital rotation \citep{patt84}. Such a spin-up consumes
the orbital angular momentum, and the MB process indirectly extracts the orbital angular
momentum from the binary system. The MB mechanism plays an important role in influencing the spin evolution of low-mass stars and the orbital evolution of binary systems with low-mass stars. At present, there exist several different MB prescriptions as follows.

\subsection{Standard MB model}
Based on the strong correlation between equatorial rotation velocity and age in low-mass main sequence (MS) stars \citep{skum72}, \cite{rapp83} derived the standard MB prescription. The rate of angular-momentum loss can be written as \citep{paxt15}
\begin{equation}
	\dot{J}_{\rm Sk}=-6.82\times10^{34}\left(\frac{M_{\rm d}}{M_{\odot}}\right)\left(\frac{R_{\rm d}}{R_{\rm \odot}}\right)^{\gamma_{\rm mb}}\left(\frac{1~\rm day}{P_{\rm orb}}\right)^3~\rm dyn~cm,
\end{equation}
where $M_{\rm d}$ and $R_{\rm d}$ are the mass and the radius of the donor star, $P_{\rm orb}$ is the orbital period, $R_{\rm \odot}$ is the solar radius, and $\gamma_{\rm mb}$ is a dimensionless parameter from 0 to 4. In this work, we adopt the simplest approximation as $\gamma_{\rm mb}=4$ \citep{verb81}.

\subsection{Matt15 MB prescription}
Observations from different stars with magnetic activity as well as some theoretical models of magnetic field generation shown that the MB torque is related to the Rossby number in the convective envelope \citep{nand04,jouv10}, which is defined as
\begin{equation}
Ro=\frac{1}{\Omega\tau_{\rm c}},
\end{equation}
where $\Omega=2\pi/P_{\rm orb}$ is the angular velocity of the star, and $\tau_{\rm c}$ is
the convective turnover timescale. The magnetic activity of various stars tends to reach a "saturated" state if the Rossby number
is less than a critical value as
\begin{equation}
Ro_{\rm sat}=\frac{Ro_{\odot}}{\chi},
\end{equation}
where $Ro_{\odot}$ is the solar Rossby number with an estimated value of $\sim 2$ \citep{see16}, and $\chi$ is a dimensionless constant. Same to \cite{goss21} and \cite{amar19}, we take $\chi=14$, then $Ro_{\rm sat}=0.14$, which is consistent with the derived value of $Ro_{\rm sat}=0.13\pm0.02$ according to the observed data \citep{wrig11}.

To better account for the distributions of the spin period and magnetic activity of sun-like stars with a low mass, \cite{matt15} derived a physically motivated MB prescription, in which the stellar wind torque depends on the Rossby number, and also includes an empirically derived scaling with stellar mass and radius. According to the models given by \cite{matt15} and \cite{amar19}, the MB torque can be written as
\begin{equation}
	\dot{J}_{\rm mb}=\left\{\begin{array}{l@{\quad}l}-T_0\left(\frac{\tau_{\rm c}}{\tau_{\rm \odot,c}}\right)^p\left(\frac{\Omega}{\Omega_{\rm \odot}}\right)^{p+1},~Ro\geq Ro_{\rm sat} \strut\\
	-T_0\chi^p\left(\frac{\Omega}{\Omega_{\rm \odot}}\right),~ Ro<Ro_{\rm sat},\strut\\ \end{array}\right.
\end{equation}
where $\tau_{\rm \odot,c}$ and $\Omega_{\odot}$ are the convective turnover time and the angular velocity of the Sun, respectively. It generally takes $\tau_{\rm \odot,c} = 2.8\times10^6~\rm s$, which was obtained by evolving a $1.0~M_{\odot}$ star with an abundance of $Z=0.02$ to an age of 4.6 Gyr \citep{van19}. Adopting an orbital period of 24 days, the angular velocity of the Sun is calculated to be $\Omega_{\rm \odot} \approx 3.0\times10^{-6}~\rm s^{-1}$. In equation (4), the torque $T_{0}$ is
\begin{equation}
T_0=K\left(\frac{R_{\rm d}}{R_{\rm \odot}}\right)^{3.1}\left(\frac{M_{\rm d}}{M_{\rm \odot}}\right)^{0.5}\gamma^{-2m},
\end{equation}
where $\gamma=\sqrt{1+(\frac{v}{0.072v_{\rm crit}})^{2}}$ \citep[$v$ and $v_{\rm crit}$ are the rotation velocity and the critical
rotation velocity of the star, respectively;][]{matt12}. The constants $K, m, p,$ and $\chi$ in the above equations are free parameters, which can be calibrated according to the observed data. In this work, we take a choice same to \cite{goss21} as $K=1.4\times10^{30}~{\rm g\,cm^{2}s^{-2}}, m=0.22, p=2.6$, and $\chi=14$.

\subsection{Garraffo18 MB prescription}
The MHD simulations indicated that the MB efficiency is tightly related to the complexity of the stellar magnetic field rather than its strength \citep{garr15,garr16,revi15}. Combined a simple MB torque prescription that is based on the Skumanich law with a function of magnetic complexity modulation, \cite{garr18} derived an MB model, in which the rate of angular momentum loss is
\begin{equation}
	\dot{J}_{\rm mb}=C\Omega^{3}\tau_{\rm c} Q_J(n),
\end{equation}
where $C=3.0\times10^{41}~\rm g\,cm^{-2}$ \citep{goss21}, $Q_J(n)=4.05e^{-1.4n}+\frac{n-1}{60Bn}$ is a modulation factor depending on the magnetic field strength $B$ and the parameter $n$ describing the
magnetic field complexity \citep{garr16}. Adopted an upper limit of $n = 7$, the modulation factor simplifies to $Q_J(n)=4.05e^{-1.4n}$ \citep{garr18}. The magnetic field complexity parameter $n$ is a function of the Rossby number and is given by \citep{garr18}
\begin{equation}
n=\frac{a}{Ro}+bRo+1.
\end{equation}
Taking $a=0.03$, and $b=0.5$, \cite{goss21} found that their solar model matches the rotation rates of Sun-like stars observed in open clusters before 4.6 Gyr.


\begin{table*}
	\begin{center}
		\centering
		\caption{Comparison between the observations of PSR J1012+5307 and our modeled results for a binary system with $M_{\rm NS,i}=1.50~M_{\rm \odot}$ and $M_{\rm d,i}=1.10~M_{\rm \odot}$ under different MB models}
		\label{tab:1}
		\begin{tabular}{c c c c c c c}
			\hline\hline\noalign{\smallskip}
			PSR J1012+5307          & Observations & Standard & Matt15 & Garraffo18 & Intermediate & CARB \\
			\hline\noalign{\smallskip}
			Initial orbital period ($\rm days$) & $\cdots$ & 2.78  & 1.09  & 1.17  & 25.2  & 8.954 \\
			Current Orbital period ($\rm days$) & 0.604    & 0.604 & 0.605 & 0.604 & 0.604 & 0.604 \\
			WD mass ($M_{\rm \odot}$)    & $0.165\pm0.015$ & 0.176 & 0.181 & 0.191 & 0.178 & 0.175 \\
			NS mass (($M_{\rm \odot}$)   & $1.72\pm0.16$   & 2.33  & 2.33  & 2.32  & 1.57  & 1.64 \\
			WD radius ($R_{\rm \odot}$)&$0.047^{+0.003}_{-0.002}$ & 0.03 & 0.022 & 0.029 & 0.049 & 0.049\\
			WD $\log g$ ($\rm cgs$)     & $6.26\pm0.04$    & 6.72 & 7.01  & 6.78  & 6.3  & 6.29 \\
			WD $T_{\rm eff}$ (K)   & $8362^{+25}_{-23}$    & 6736  & 1478  & 7251  & 12284 & 11999 \\
			Age ($\rm Gyr$)            & $\cdots$          & 25    & 93    & 43   &  9.9  &  10\\
			\hline\noalign{\smallskip}	
		\end{tabular}
	\end{center}
\end{table*}

\subsection{`Intermediate' MB}
Considering the influence of angular velocity, convective turnover time, and wind-loss rate, \cite{van19} modified the standard MB prescription to be
\begin{equation}
	\dot{J}_{\rm boost}=\dot{J}_{\rm Sk}\left(\frac{\Omega}{\Omega_{\rm \odot}}\right)^\beta \left(\frac{\tau_{\rm c}}{\tau_{\rm \odot,c}}\right)^\xi \left(\frac{\dot{M}_{\rm w}}{\dot{M}_{\rm \odot,w}}\right)^\alpha,
\end{equation}
where $\dot{M}_{\rm w}$ is the wind-loss rate of the companion star. Furthermore, the solar wind-loss rate is taken to be $\dot{M}_{\rm \odot,w}=2.5\times10^{-14}~M_{\odot}\rm yr^{-1}$ \citep{carr06}.

In equation (8), three power indices $\beta,\xi,\alpha$ are used to determine the influence of the surface magnetic field, convective turnover time, and wind-loss rate on the rate of angular momentum loss, respectively. When $\beta=0,\xi=0,\alpha=0$, equation (8) is the same as the standard MB prescription proposed by \cite{rapp83}, that is 'default' Skumanich MB case \citep{van19}. When $\beta=0,\xi=2,\alpha=0$, the scale of convective turnover time is taken into account, i.e. $\dot{J}_{\rm boost}=\dot{J}_{\rm Sk}(\tau_{\rm c}/\tau_{\rm \odot,c})^{2}$, which was called the 'convection-boosted' MB case. The case with $\beta=0,\xi=2,\alpha=1$ considers two scaling terms including both the convective turnover time and wind-loss rate, i.e. 'intermediate' MB case, in which the additional wind scaling term is linear in wind mass-loss rate. Based on a detailed one-dimensional stellar evolution model, \cite{van19} found that the `Intermediate' MB prescription can produce the largest number of observed transient LMXBs, as well as super-Eddington mass-transfer rate observed in the LMXB 2A1822-371 \citep{van19}. The fourth case with $\beta=2,\xi=4,\alpha=1$ is the "wind-boosted" MB case, in which the three scaling terms of magnetic field strength, convective turnover time, and wind mass loss rate are included. \cite{van19} found that the stellar evolution simulations using the "wind-boosted" MB prescription cannot reproduce the majority of the observed persistent
or transient X-ray binaries. Therefore, in this work we adopt the 'Intermediate' MB case to model the formation and evolution of LMXBs, and attempt to reproduce the observational properties of PSR J1012+5307.

\subsection{Convection- and rotation-boosted MB}
To solve the discrepancies between the observed mass-transfer rates and the theoretical models in LMXBs, \cite{vani19} proposed a modified MB prescription, that is, convection- and rotation-boosted (CARB) MB. This new MB model considered the influence of the
donor-star's rotation on the wind's velocity \citep{matt12,revi15} and the surface magnetic field \citep{park71,noye84,ivan06,van19}, as well as the effects of the
donor star's convective eddy turnover timescale on the rate of angular momentum loss. In the CARB MB prescription, the rate of angular momentum loss can be expressed as \citep{vani19}
\begin{equation}
    \begin{aligned}
	\dot{J}_{\rm MB}=-\frac{2}{3}\dot{M}^{-1/3}_{\rm w}R^{14/3}_{\rm d}(\nu^2_{\rm esc}+2\Omega^2R^2_{\rm d}/K^2_2)^{-2/3}\\
	\times\Omega_{\rm \odot}B^{8/3}_{\rm \odot} \left(\frac{\Omega}{\Omega_{\rm \odot}}\right)^{11/3}\left(\frac{\rm \tau_{\rm c}}{\tau_{\rm \odot,c}}\right)^{8/3}
    \end{aligned}
\end{equation}
where $\nu_{\rm esc}$ is the surface escape velocity of the donor star, $B_{\rm \odot} = 1 \rm G$ is the surface magnetic field strength of the Sun \citep{van19}, $K_2$ determines the limit where high rotation rate starts to produce small angular-momentum-loss rate, and a grid of simulations given by \cite{revi15} derived a constant $K_2$ as $K_2=0.07$.

\begin{figure}
	\centering
	\includegraphics[width=1.1\linewidth,trim={0 0 0 0},clip]{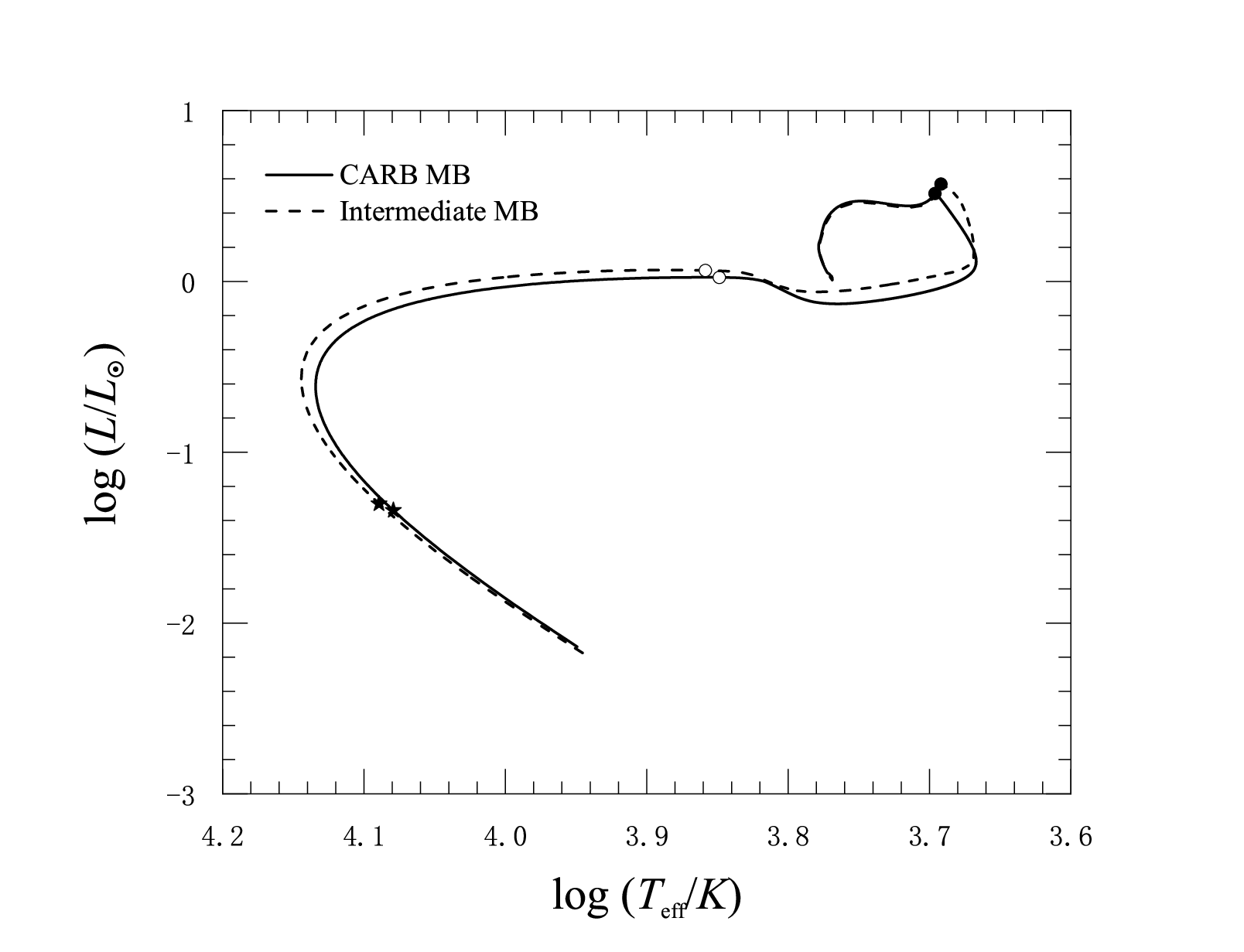}
	\\	
	\centering
	\includegraphics[width=1.1\linewidth,trim={0 0 0 0},clip]{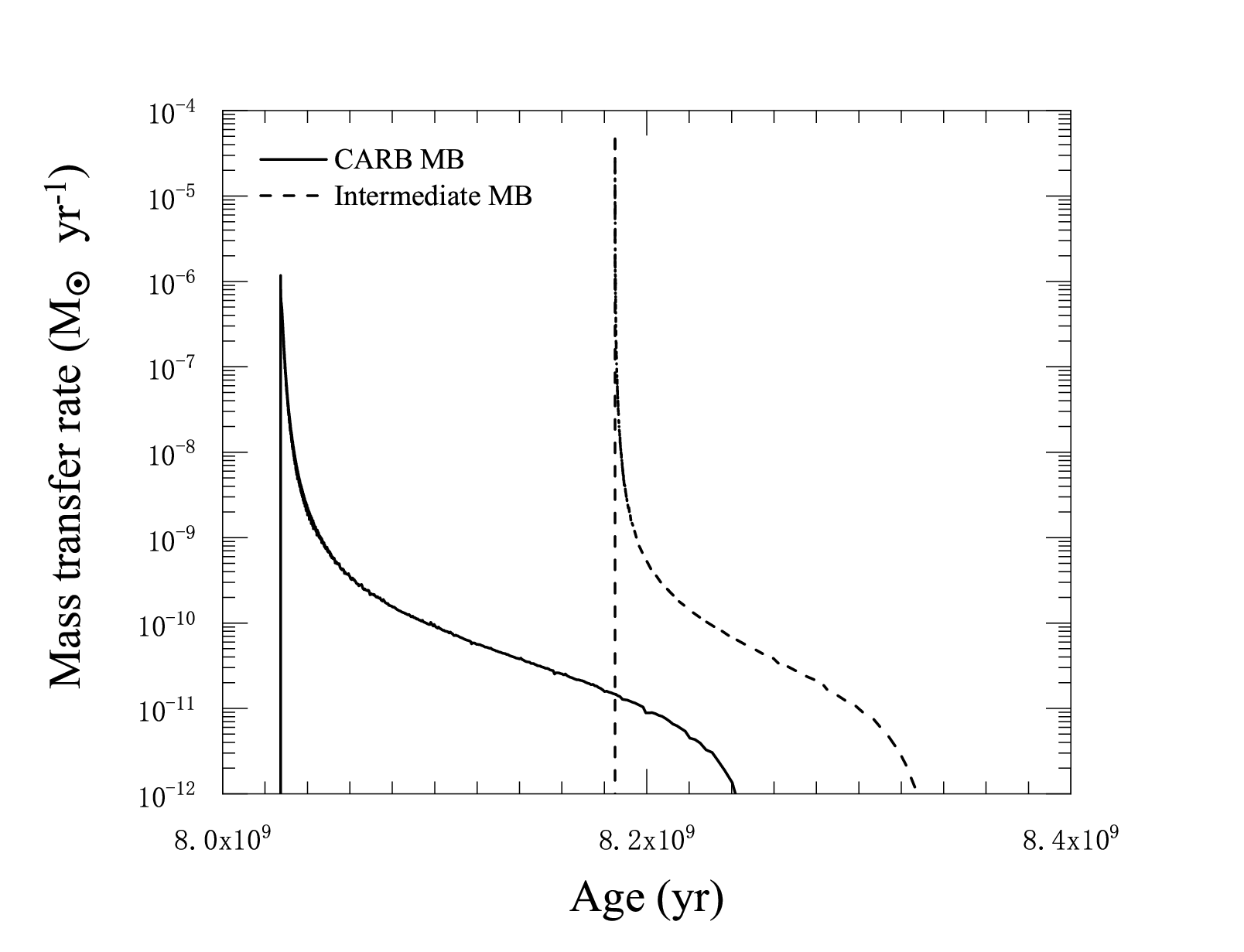}
	\caption{Evolution of NS-MS binaries in H-R diagram (top panel) and mass-transfer rate vs. stellar age diagram (bottom panel) for the best models of PSR J1012+5307. The solid curve represents the case of the CARB MB model, in which the initial parameters are $M_{\rm NS,i}=1.50~M_{\rm \odot}$, $M_{\rm d,i}=1.10~M_{\rm \odot}$, and $P_{\rm orb,i}=8.954~{\rm days}$. The dashed curves the case of the 'Intermediate' MB model, in which $M_{\rm NS,i}=1.50~M_{\rm \odot}$, $M_{\rm d,i}=1.10~M_{\rm \odot}$, and $P_{\rm orb,i}=25.20~{\rm days}$. The solid circles and open circles in the upper panel mark the onset and the termination of mass transfer, respectively. The solid stars in the upper panel represent the positions that can well match the observed WD mass and orbital period of PSR J1012+5307 (see also Table 1).}
	\label{fig:orbmass}
\end{figure}

\begin{figure*}
	\subfigure{
		\centering
		\includegraphics[width=0.5\linewidth,trim={0 0 0 0},clip]{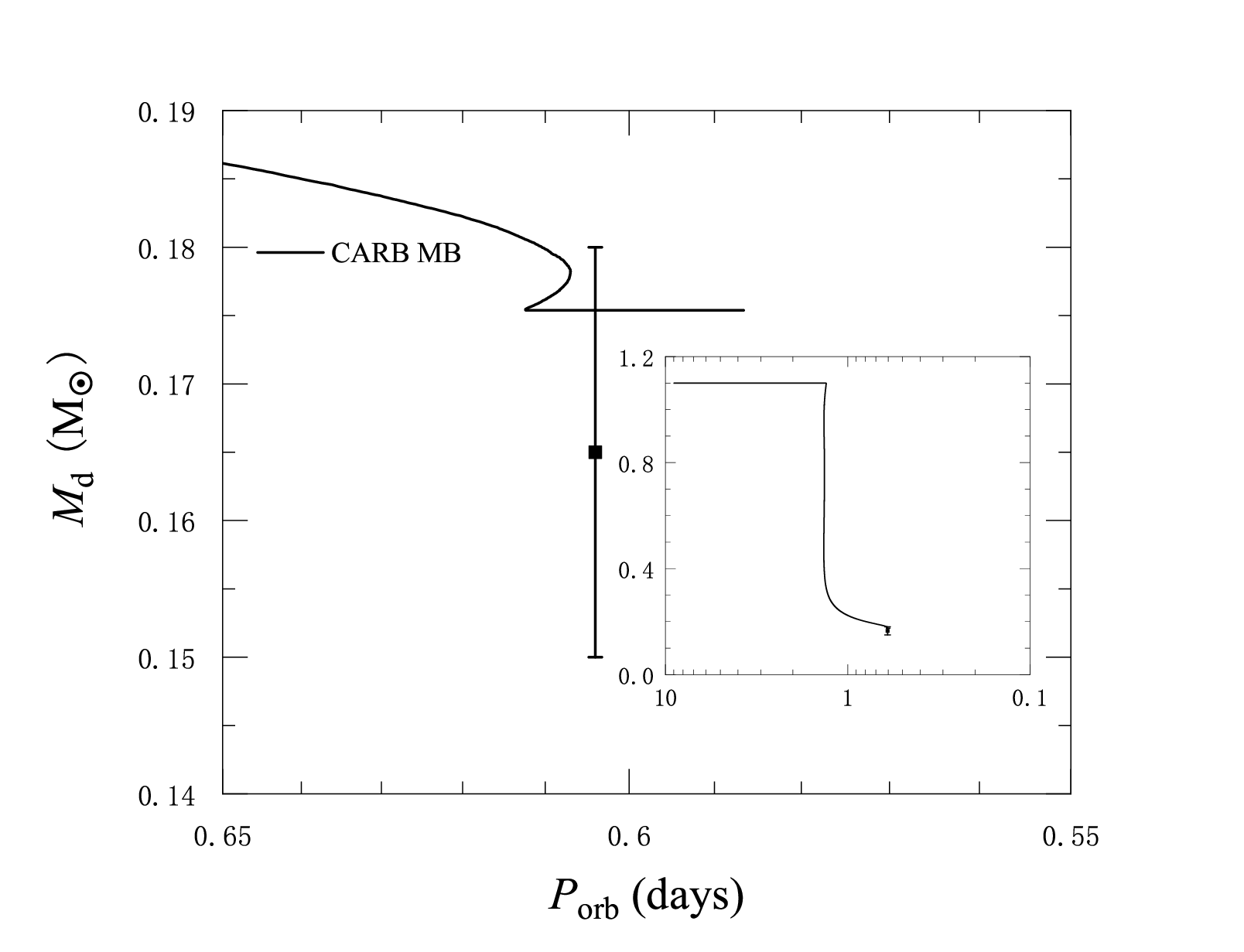}
	}
	\subfigure{
		\centering
		\includegraphics[width=0.5\linewidth,trim={0 0 0 0},clip]{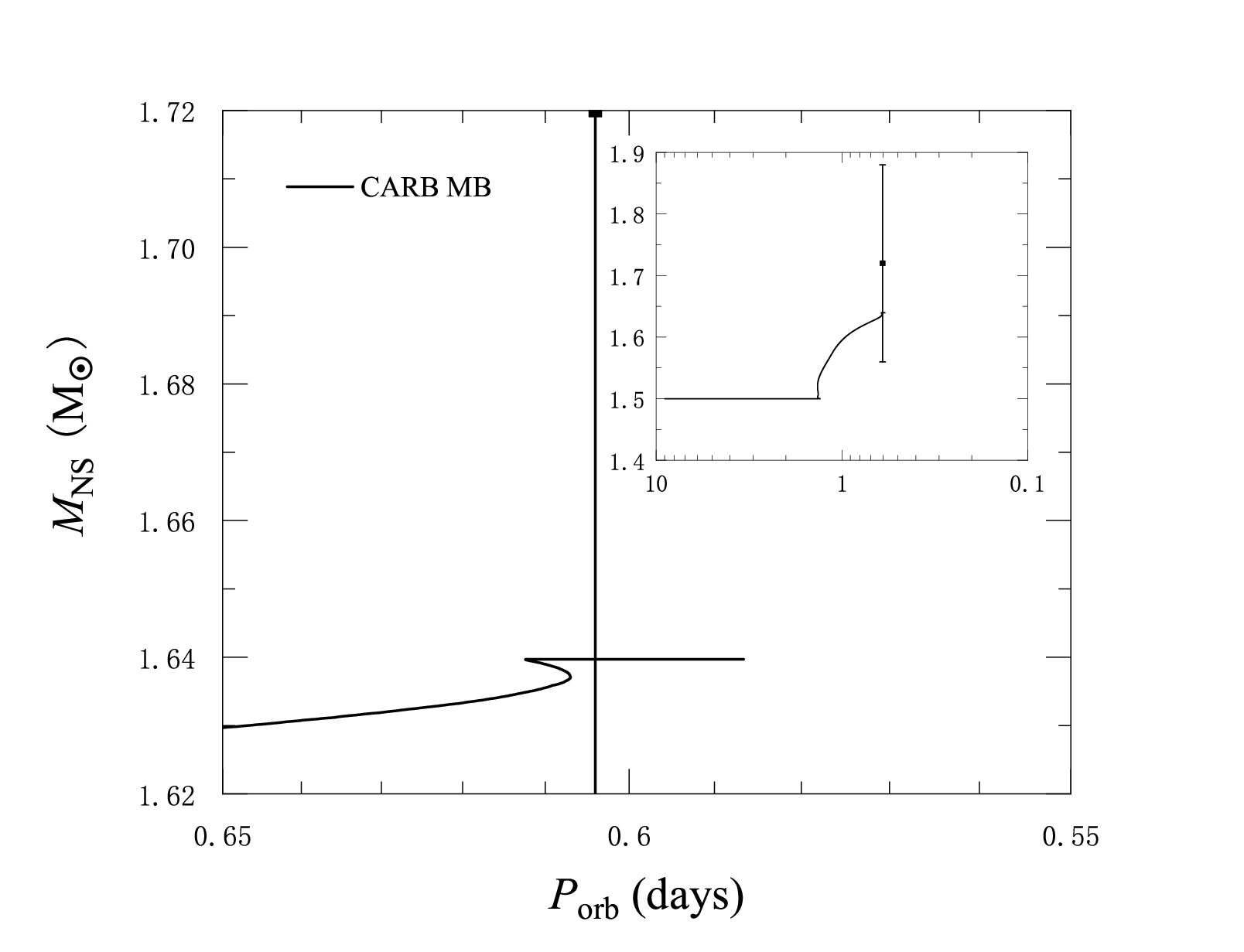}
	}
	\\
	\subfigure{
		\centering
		\includegraphics[width=0.5\linewidth,trim={0 0 0 0},clip]{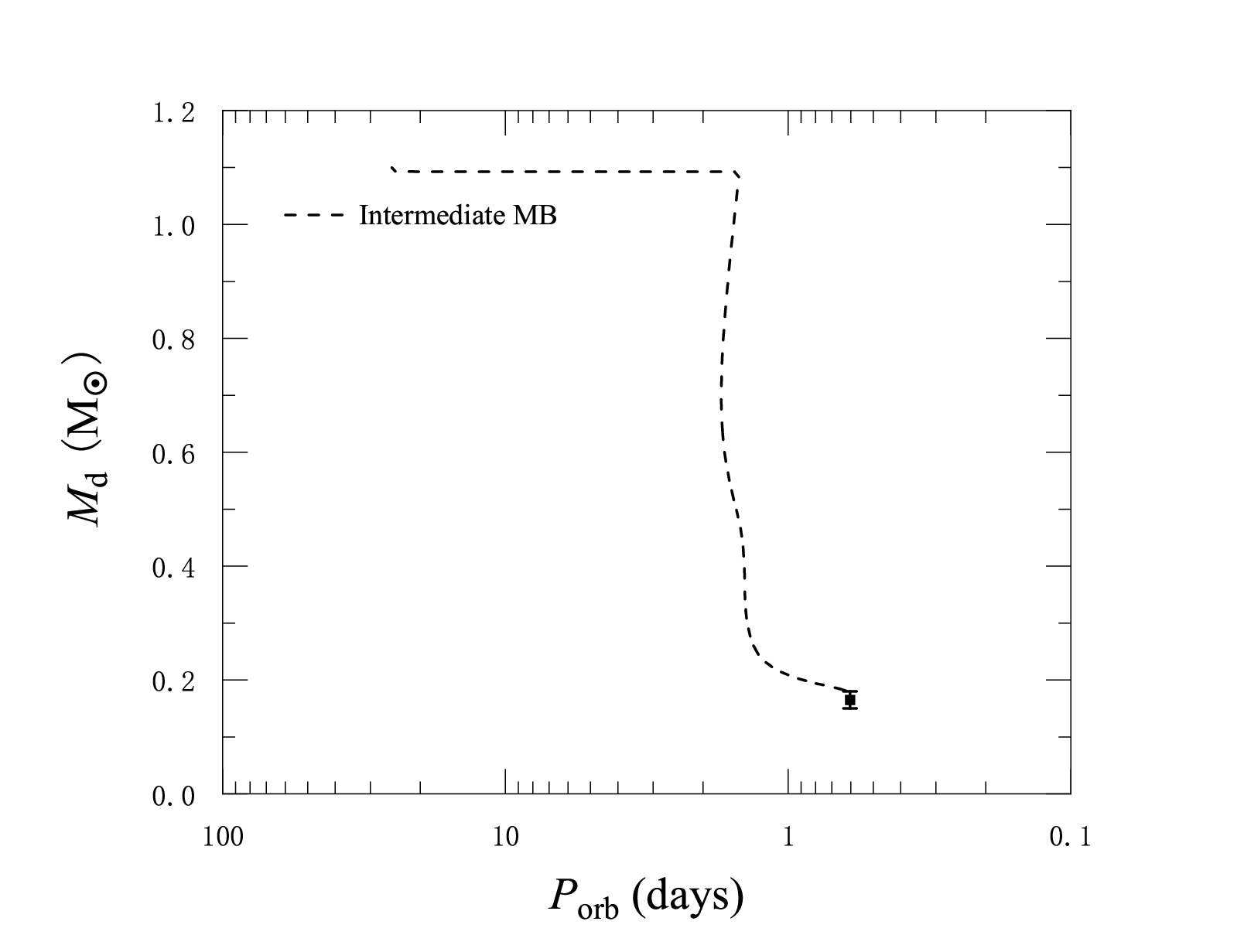}
	}
	\subfigure{
		\centering
		\includegraphics[width=0.5\linewidth,trim={0 0 0 0},clip]{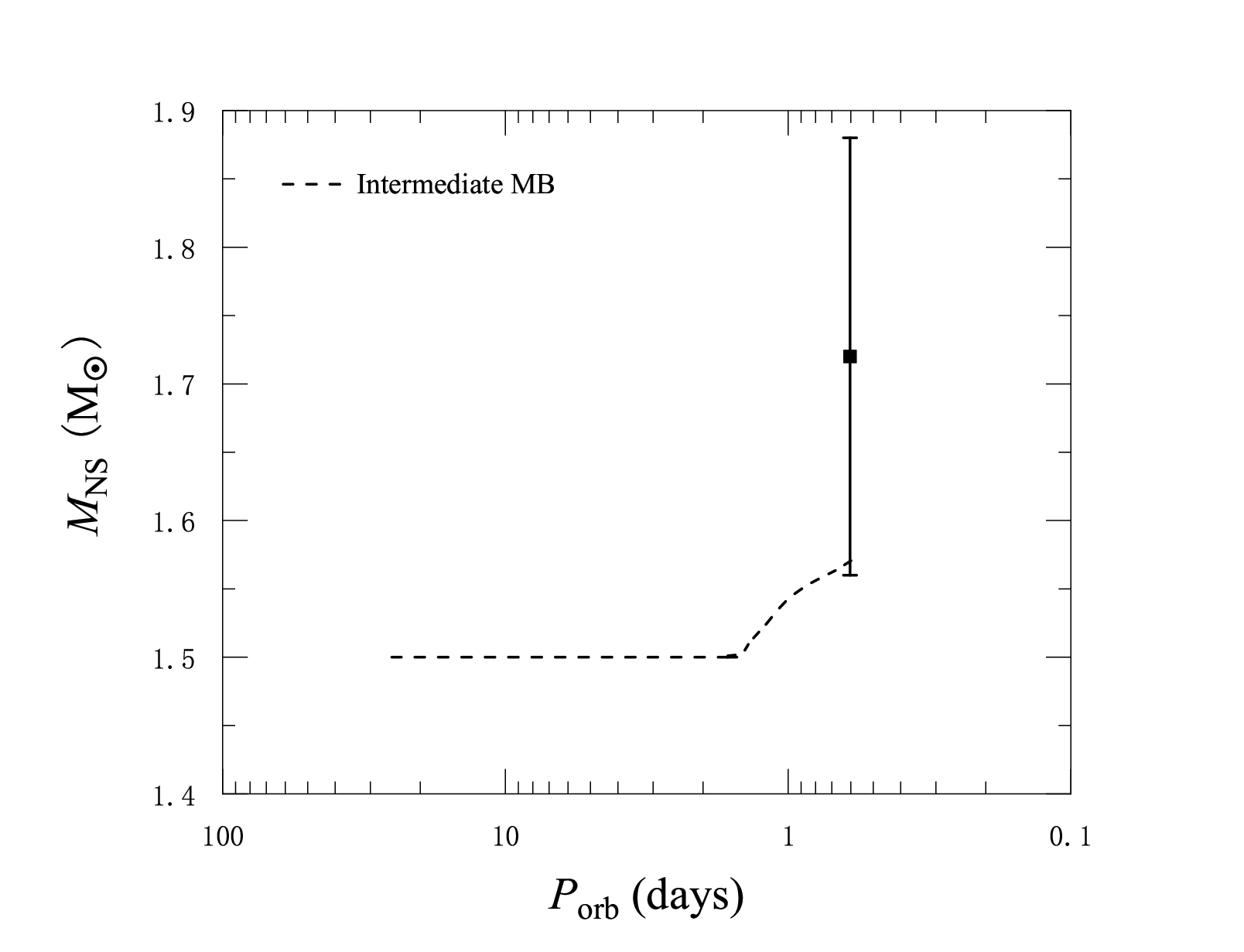}
	}
	\caption{Evolution of the donor-star mass (left panels) and the NS mass (right panels) with the orbital period for our best models reproducing the observed parameters of PSR J1012+5307. The initial parameters of the NS binary are the same as in Figure 1. The solid curves and dashed curves correspond to the CARB MB and the 'Intermediate' MB models, respectively. The solid squares with error bars represent the observed parameters of PSR J1012+5307.}
	\label{fig:orbmass}
\end{figure*}

\section{Stellar evolution simulations}
\subsection{Evolutionary code}
To understand the origin of PSR J1012+5307 with an ELM WD, we perform a detailed model for the evolution of a binary system consisting of a NS and a low-mass MS companion. The open-source software instrument Modules for Experiments in Stellar
Astrophysics \citep[MESA, version r12115,][]{paxt11,paxt13,paxt15,paxt18,paxt19} is used to calculate the nuclear synthesis of the donor star and the orbital evolution of the binary. The MS companion star is initially arranged in solar chemical composition, i.e., $X=0.70, Y=0.28, Z=0.02$. In the calculation, the NS is considered a point mass, and the orbit
of the binary is assumed to be circular and tidally synchronized.

Once the donor star fills the Roche lobe, its surface H-rich envelope is transferred to the NS at a rate of $\dot{M}_{\rm tr}$.
During the accretion, the mass-growth rate of the NS is limited by the Eddington accretion rate ($\dot{M}_{\rm Edd}=1.5\times 10^{-8}~M_{\odot}\,\rm yr^{-1}$), i.e. $\dot{M}_{\rm NS}={\rm min}(\dot{M}_{\rm tr},\dot{M}_{\rm Edd})$. The excess material in unit time ($\dot{M}_{\rm tr}-\dot{M}_{\rm NS}$) is re-emitted as an isotropic fast wind in the vicinity of the NS, carrying away the specific orbital angular momentum of the NS.

In the calculation, we consider four angular-momentum-loss mechanisms including gravitational wave radiation, MB, mass loss, and angular momentum loss due to spin-orbit coupling. For each system, we adopt different MB models described in Section 2 to model the evolution of the NS binary. The MB mechanism is arranged to operate if the donor star possesses both a convective envelope and a radiative core \citep{paxt15}.

\subsection{Initial values}
To obtain a model that can reproduce the observed properties of PSR J1012+5307, we calculate a grid of binary evolution for different initial donor-star masses and initial orbital periods. According to the currently measured mass $M_{\rm NS}=1.72\pm0.16~M_{\rm \odot}$ of the NS, we take the initial NS mass to be $M_{\rm NS,i}=1.50~M_{\rm \odot}$. It implies that the NS of PSR J1012+5307 was born massive. Similar phenomenons were also proposed to interpret the formation of PSR
J1614-2230 and PSR J1640+2224, in which the initial NS masses were inferred to be
born with masses of $\sim1.95~M_{\odot}$ \citep{taur11} and $>2.0~M_{\odot}$ \citep{deng20}. The donor-star mass ranges from 1.0 to $1.5~M_{\odot}$ in steps of $0.1~M_{\odot}$, and the initial orbital period ranges from 0.4 to 100 days in steps of $\bigtriangleup {\rm log}(P_{\rm orb}/{\rm days})=0.2$. In the grid, we search for a model whose calculated parameters at $P_{\rm orb}=12.5~\rm hours$ are relatively close to the properties of PSR J1012+5307. Subsequently, we fine-tune the initial orbital period to find the best model that can reproduce the observed properties of PSR J1012+5307.

\subsection{Simulated Results}
Our simulation finds that a binary system with a $M_{\rm d,i}=1.10~M_{\rm \odot}$ and different initial orbital periods can reproduce the detected orbital period and WD mass for different MB models. The comparison of the best models for the five different MB prescriptions and the observed data are summarized in Table 1. To evolve to the observed orbital period and WD mass of PSR J1012+5307, the NS binaries under the standard,  Matt15, and Garraffo18 MB mechanisms spend long evolutionary timescales of 25, 93, and 43 Gyr, which are much longer than the Hubble time. This is because these MB mechanisms are inefficient in driving the evolution of LMXBs (see also section 4.1). Meanwhile, these three MB models can not reproduce the NS mass, the radius, the surface gravity, and effective temperature of the WD. Therefore, these three MB mechanisms are highly unlikely to drive NS LMXBs to evolve into PSR J1012+5307. The CARB MB and the 'Intermediate' MB models can well reproduce the observed properties of PSR J1012+5307 except for the effective temperature for the initial orbital period $P_{\rm orb}=8.954$ and 25.2 days, respectively. Therefore, in this section we only compare the results of the two models by the CARB MB and `Intermediate' MB prescriptions with the observations of PSR J1012+5307.

To show the evolutionary history and fates of the donor stars, in the top panel of Figure 1, we plot their evolutionary tracks in an H-R diagram. The evolutionary tracks of both MB models are very similar. At $L\sim L_{\odot}$ and $T_{\rm eff}\sim7000~\rm K$, the donor star is striped the H-rich envelope, and the remaining He core decouples from the Roche lobe. Subsequently, the He core starts contraction and cooling process and evolves into an ELM WD. At the present orbital period of PSR J1012 + 5307, the luminosity and effective temperature of the ELM WD are ${\rm log}(L/L_{\odot})\sim-1.3$ and $T_{\rm eff}\sim12000~\rm K$, respectively.

The bottom panel of Figure 1 shows the evolution of the mass-transfer rate. For the CARB MB model, the donor star fills the Roche lobe at $t\approx8.0\times10^{9}~{\rm yr}$, and starts to transfer the surface H-rich material onto the NS at a rate of $\dot{M}_{\rm tr}\sim10^{-6}~M_{\rm \odot}~{\rm yr^{-1}}$.  For the 'Intermediate' MB model, the stellar age started the mass transfer and the maximum mass-transfer rate are $t\approx8.2\times10^{9}~{\rm yr}$ and $\dot{M}_{\rm tr}\sim10^{-4}~M_{\rm \odot}~{\rm yr^{-1}}$, respectively. Because the mass is transferred from the less massive donor star to the more massive NS at a high rate, the orbit of the binary continuously widens in the early mass-transfer stage. When the mass-transfer rate gradually decreases, the MB mechanism dominates the orbital evolution, and the orbital period of the binary continuously decreases. Once the H-rich envelope is depleted, the donor star decouples from its Roche lobe. The two mass-transfer timescale for two MB models are $\sim0.2~\rm Gyr$. Since the orbital periods are too long ($0.6~\rm days$) when the systems evolve into detached binaries, the gravitational-wave radiation can not drive the second mass transfer in the Hubble time, and the systems can not evolve toward ultracompact X-ray binaries \citep[the second mass transfer can achieve if the initial orbital period in the detached stage is $\sim0.3~\rm days$, see also][]{chen20}.

Figure 2 displays the evolution of the WD masses and NS masses with orbital periods of our best models for PSR J1012+5307. Before the donor stars fill their Roche lobes, two MB mechanisms drive the orbital periods to decrease to $\sim 1.5~\rm days$. It seems that the angular-momentum-loss rate of the `Intermediate' MB mechanism is much higher than that of the CARB MB model because the initial orbital period is 25.2 days for the former case (such an initial orbital period is much longer than 8.954 days for the latter). Once the mass transfer initiates, the H-rich material on the surface of the donor stars is transferred from the less massive donor star to the more massive NS, which tends to cause the expansion of the orbit. However, the angular-momentum loss due to MB produces a negative orbital-period derivative. Therefore, the evolutionary fate of the orbit depends on the competition between the mass transfer and the MB mechanism. In the early stage of the mass transfer, an ultra-high mass transfer rate dominates the orbital evolution, resulting in an orbital expansion. With the decline of the mass-transfer rate, the MB mechanism dominates the orbital evolution and causes the orbit to gradually shrink.

In the CARB MB case, the orbital period decreases to the minimum of 0.607 days, then increases. When the orbital period is $0.612~{\rm days}$, the donor star decouples from the Roche lobe, and the system evolves into a detached binary. Subsequently, the donor star with a mass of $0.175~M_{\odot}$ first evolves into an ELM He WD via a $\sim1$ Gyr contraction stage, and then begins a cooling track \citep{istr14b}. In this stage, the gravitation radiation drives the orbital period to decrease to the observed period of 0.604 days of PSR J1012+5307. \cite{laza09} derived the intrinsic change rate of the orbital period of PSR J1012+5307 to be $\dot{P}_{\rm int}=(-1.5\pm1.5)\times10^{-14}~\rm s\,s^{-1}$, which agrees well with the contribution ($\dot{P}_{\rm gw}=(-1.1\pm0.2)\times10^{-14}~\rm s\,s^{-1}$) due to gravitation radiation. This provides direct evidence that this source is in the detached stage, which is consistent with our simulated results.

Our simulated WD mass is consistent with the WD mass range of $0.167-0.175~M_{\odot}$ for an orbital period in the range of $0.412-0.628$ days when the gravitational settling, thermal and chemical diffusion, and rotational mixing of donor stars with $Z=0.02$ was included \citep[see also Table A.1 of][]{istr16}. Because a relatively high angular-momentum-loss rate, the 'Intermediate' MB mechanism drives the orbital period to decrease to 0.61 days (the orbital period firstly decreases to 1.49 days, then increases to 1.72 days, finally decreases to 0.61 days), at which the donor star with a mass of $0.178~M_{\odot}$ decouples from its Roche lobe. Subsequent evolution is the same as the CARB case. The final NS masses are $1.64~M_{\odot}$ and $1.57~M_{\odot}$ for the CARB and 'Intermediate' MB prescriptions, respectively. Our simulated results indicate that both the CARB MB and the 'Intermediate' MB models can reproduce the observed NS-mass and WD-mass at the current orbital period of PSR J1012+5307.

In the left upper panel of Figure 3, we plot the evolution of the orbital period with stellar age for our best models reproduced the formation of PSR J1012+5307. The NS binaries can evolve into the current orbital period ($P_{\rm orb}=0.604~\rm days$) of  PSR J1012+5307 for the CARB and the `Intermediate' MB models at the stellar age $t=10~{\rm Gyr}$ and $9.9~{\rm Gyr}$, respectively. The CARB MB case initiates mass transfer earlier than the 'Intermediate' MB case. However, the mass-transfer timescale of the 'Intermediate' MB case is shorter than that of the CARB MB case because of a high mass-transfer rate for the former. The evolution of the donor-star effective temperature and radius with orbital period is displayed in the right upper right panel, and the left bottom panel, respectively. In the early stage, the effective temperature decreases due to the expansion of the donor star. After the mass transfer ceases, the remaining He core gradually contract due to gravity, resulting a rapid climb of the effective temperature. Once the He core evolves into a He WD, it would begin a cooling process, and the effective temperatures at the current orbital period are 11999, and 12284 K for the CARB and the 'Intermediate' MB models, respectively. These two temperatures are higher than the observed effective temperature of $8362^{+25}_{-23}~\rm K$.

The right bottom panel of Figure 3 depicts the evolution of the surface gravity of the donor star with the orbital period. It emerges three "knee" features in the evolutionary tracks of the donor-star radius and the donor-star surface gravity. The first "knee" arises from the beginning of a mass transfer. In the CARB MB and the `Intermediate' MB cases, the donor star fills its Roche lobe and triggers a mass transfer at $P_{\rm orb}=1.31~{\rm days}$, and $1.50~{\rm days}$, respectively. A change of the mechanism dominating the orbital evolution results in the second "knee", at which the dominant mechanism changes from the mass transfer to the angular-momentum loss due to the MB. The terminal of the mass transfer is responsible for the third "knee". The evolutionary tracks of the donor-star radius and its surface gravity are very similar, which arises from their anti-correlative relation. At the present orbital period, the radius and the surface gravity of the WD for the CARB ('Intermediate') MB prescription are ${\rm log}g=6.29~(6.3)$, and $R_{\rm wd}=0.049~(0.049)~R_{\odot}$, respectively, which are well in agreement with the observations of  PSR J1012+5307.

\begin{figure*}
	\subfigure{
		\centering
		\includegraphics[width=0.5\linewidth,trim={0 0 0 0},clip]{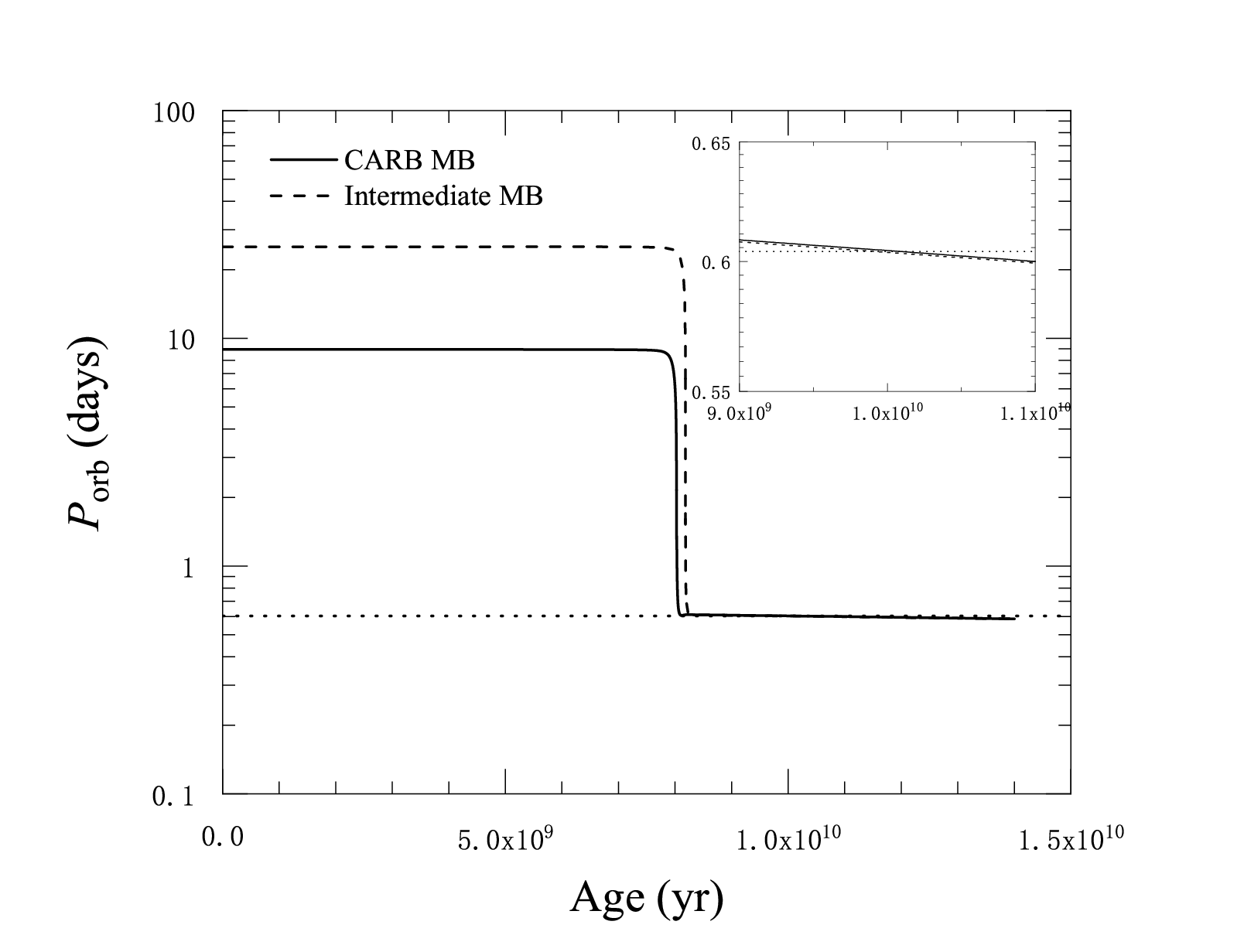}
	}
	\subfigure{
		\centering
		\includegraphics[width=0.5\linewidth,trim={0 0 0 0},clip]{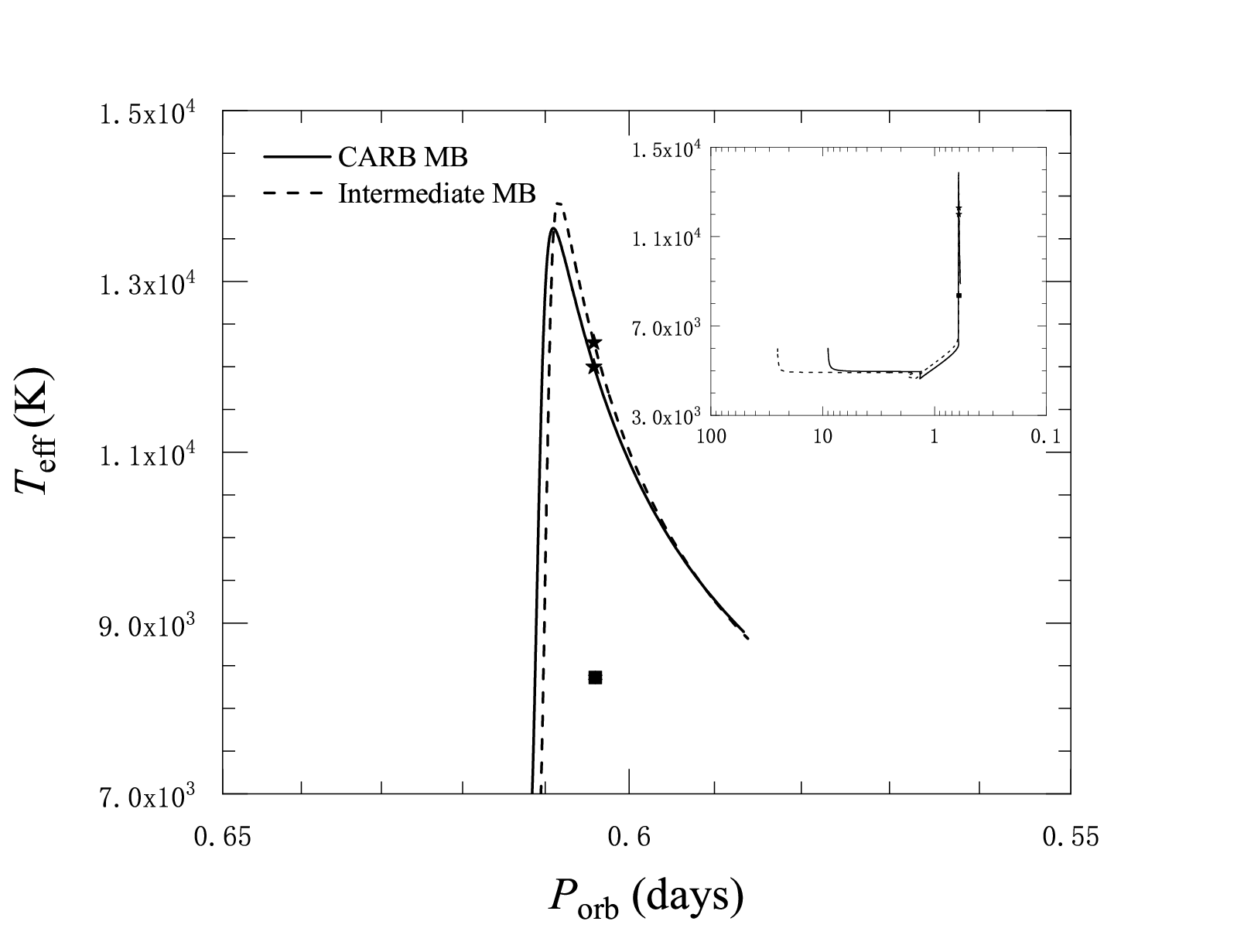}
	}
	\\
	\subfigure{
		\centering
		\includegraphics[width=0.5\linewidth,trim={0 0 0 0},clip]{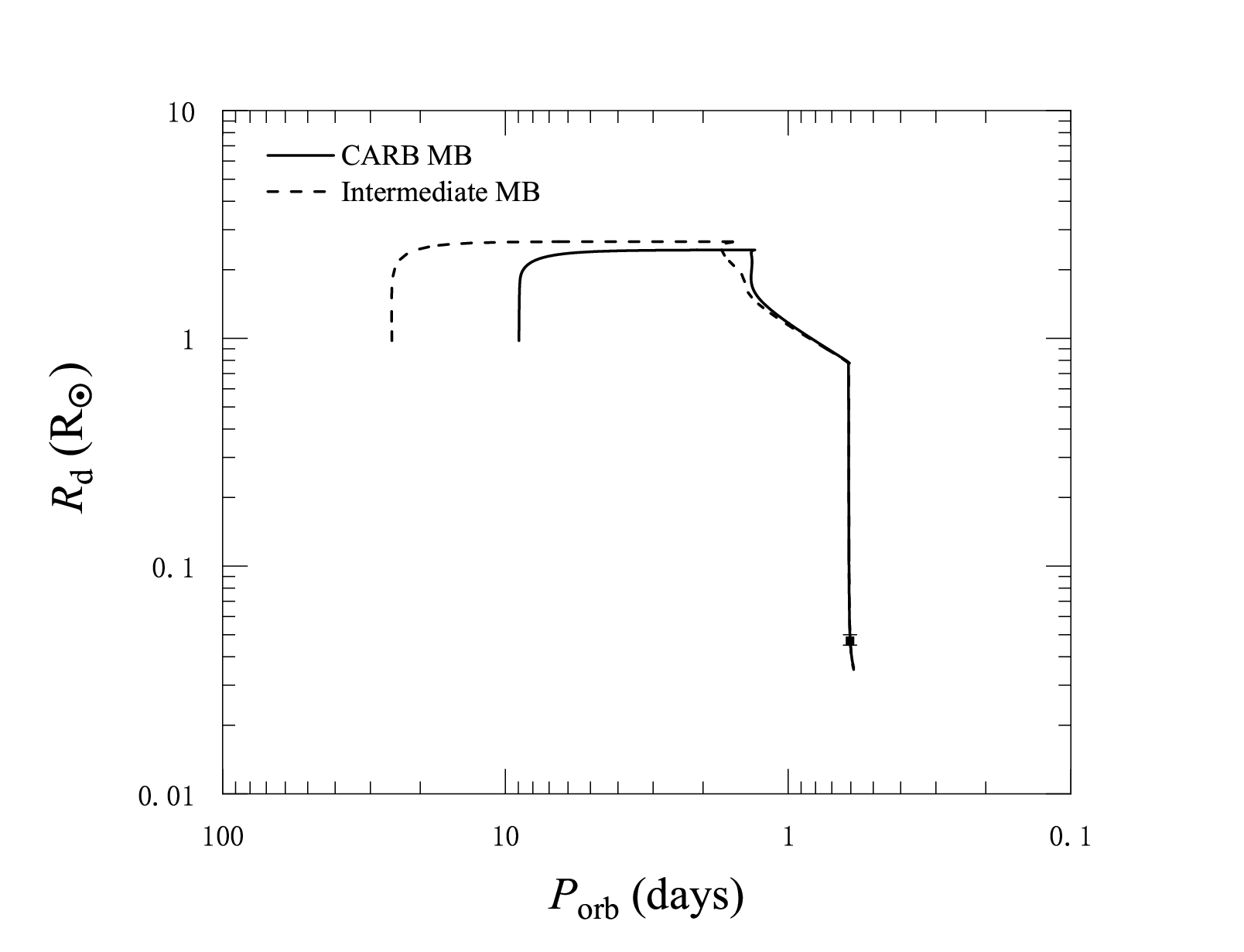}
	}
	\subfigure{
		\centering
		\includegraphics[width=0.5\linewidth,trim={0 0 0 0},clip]{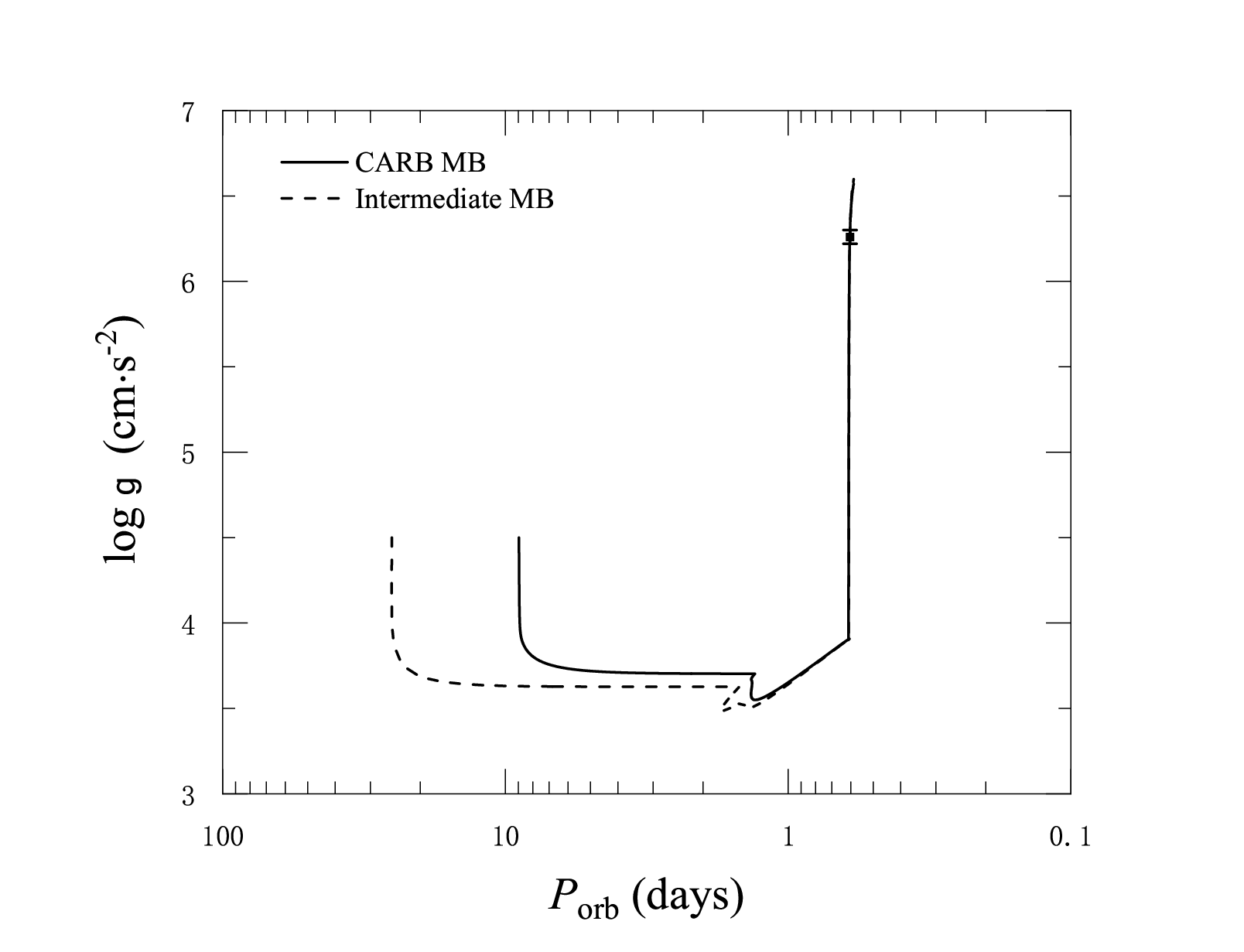}
	}
	\caption{Evolution of the orbital period with the stellar age (left upper panel), and evolution of the donor-star effective temperature (right upper panel), the donor-star radius (left bottom panel), and the WD surface gravity (right bottom panel) with the orbital period for our best models reproducing the observed parameters of PSR J1012+5307. The initial parameters of the NS binary are the same as in Figure 1. The solid and dashed curves correspond to the CARB MB and the 'Intermediate' MB models, respectively. The solid squares with error bars represent the observed parameters of PSR J1012+5307. The horizontal short dotted line indicates the observed orbital period of PSR J1012+5307. The solid stars in the upper right panel denote our simulated WD effective temperature, at which the simulated orbital period matches the observation of PSR J1012+5307.}
	\label{fig:orbmass}
\end{figure*}

\section{Discussion}
\subsection{Comparison between different MB prescriptions}
Our detailed binary evolution models show that the CARB and `Intermediate' MB prescriptions can reproduce all observed parameters of PSR J1012+5307 except for the effective temperature of the WD in the Hubble time. The other three MB prescriptions can neither produce the radius, surface gravity, and effective temperature of the WD nor evolve to the present orbital period and the WD mass in the Hubble time. This discrepancy should originate from the different rates of angular momentum loss of different MB prescriptions.

Figure 4 displays the evolution of the rate of angular momentum loss of different MB prescriptions for a NS binary with $M_{\rm NS,i}=1.50~M_{\rm \odot}$, $M_{\rm d,i}=1.10~M_{\rm \odot}$, and $P_{\rm orb,i}=1.1~{\rm days}$. Both the CARB and `Intermediate' MB prescriptions possess much higher rates of angular momentum loss. However, the rates of angular momentum loss for the Matt15 and Garraffo18 MB prescriptions are approximately three orders of magnitude smaller than those for the CARB and `Intermediate' MB models and are about two orders of magnitude smaller than that for the standard MB model.

The simulated mass-transfer rates in low-mass X-ray binaries (LMXBs) were early found too low. Statistical analyses found that the birthrate of low-mass binary radio pulsars is an order of magnitude lower than that of their progenitor LMXBs, and this discrepancy implied that the actual mass transfer rates in LMXBs are much higher than those predicted from stellar evolution models \citep{kulk88}. Some works already noticed that the observed mass-transfer rates in some LMXBs with short orbital periods are at least an order of magnitude higher than the simulated values under the standard MB model \citep{pods02,pfah03,shao15,pavl16}. \cite{van19} also showed that the standard MB prescription predicts mass transfer rates an order of magnitude too low to match most observed persistent LMXBs.

Considering a scaling of the magnetic field with the convective turnover time, and a scaling of MB with the wind-loss rate, \cite{van19} modified the MB law, and found that the 'Intermediate' MB prescription can produce the largest number
of observed transient LMXBs. Including the influence of the stellar rotation on the stellar wind velocity, the influences of the stellar convective-turnover timescale, and the stellar rotation on the surface magnetic field, \cite{vani19} derived the CARB MB law. The detailed binary evolution models with the CARB MB prescription can reproduce the mass-transfer rates and orbital periods of all observed persistent NS LMXBs with the detected mass ratio in the Galaxy \citep{vani19}. Adopting the CARB MB model, the simulated results of detailed binary evolution can match the observed parameters of all persistent NS LMXBs and binary pulsars \citep{deng21}. Recently, it found that the CARB MB model can produce a super-Eddington mass transfer in persistent NS LMXBs with low-mass ($\sim0.5~M_{\odot}$) donor stars \citep{vani19,vani21}, interpreting the formation of 2A 1822-371 with a rapid orbital expansion \citep{wei23}.

The Matt15 MB prescription can account for some broad features of the stellar spin distribution observed in the Kepler field \citep{matt15}. However, the model can not reproduce those slow rotators with masses below $0.5~M_{\rm \odot}$ in the Kepler field. Similarly, all of the stars with masses of $0.3-0.8~M_{\rm \odot}$ predicted by the Matt15 MB law rotate too quickly to match the observation that most stars in this mass range are slow rotators \citep{doug17,goss21}. This implies that the Matt15 MB prescription may underestimate the rate of angular momentum loss, which is consistent with our simulations.

The Garraffo18 MB prescription can reproduce the bimodal distribution of both slow and fast rotators observed in
young open clusters \citep{garr18}. However, this MB model predicted that those stars in the mass
range of $0.3-0.8~M_{\rm \odot}$ tend to spin too slowly \citep{goss21}. Our simulation also indicate that the angular
momentum loss of Garraffo18 MB prescription is more efficient than that in the case of Matt15 when the stellar mass is less than $0.55~M_{\odot}$ (see also our Figure 4).

For a binary system with $M_{\rm NS}=1.4~M_{\odot}$, $M_{\rm d}=1.0~M_{\odot}$, and $P_{\rm orb}=1.0~\rm days$, the orbital angular momentum $J_{\rm orb}=\frac{M_{\rm NS}M_{\rm d}}{M_{\rm NS}+M_{\rm d}}a^{2}\Omega=1.3\times10^{52}~\rm g\,cm^{2}s^{-1}$, however the spin angular momentum of the donor star is $J_{\rm s}=0.4M_{\rm d}R_{\rm d}^{2}\Omega=2.8\times10^{50}~\rm g\,cm^{2}s^{-1}$. The total orbital angular momentum of the binary is approximately two orders of magnitude greater than the spin angular momentum of the donor star. Therefore, both the Matt15 and Garraffo18 MB prescriptions should be efficient in the influence of the spin evolution of single stars, while they are too weak to affect the evolution of LMXBs.

\subsection{Comparison with previous works  }
Based on the evolutionary tracks for low-mass WDs with a mass in the range of $0.179-0.414~M_{\odot}$, \cite{drie98} determined the WD mass of PSR J1012+5307 to be $0.19\pm0.1~M_{\odot}$. However, they adopted a surface gravity of ${\rm log}g=6.75\pm0.07$ \citep{kerk96}, and did not take into account the evolution of the orbital period. In the Hubble time, the standard MB model can account for the observed orbital period and a WD mass of $0.176~M_{\odot}$  of PSR J1012+5307 \citep{mata20}. However, they adopted a metallicity of $Z=0.014$ (a low metallicity can result in a short evolutionary timescale), and the WD has also a high effective temperature of 15500 K when its radius $R_{\rm WD}=0.05~R_{\odot}$ \citep{mata20}. A WD with a mass of $0.165~M_{\odot}$ can account for the observed radius and effective temperature, however, the orbital period is smaller than 0.3 days when the donor star evolves into this stage of the WD \citep{mata20}. \cite{alth13} performed a detailed grid of evolutionary sequences for He-core WDs for an entire mass range ($0.15-0.35~M_{\odot}$) and an initial donor-star abundance of $Z=0.01$, in which the WD of PSR J1012+5307 has an effective temperature of $T_{\rm eff}=8670\pm300~\rm K$, a surface gravity of ${\rm log}g=6.34\pm0.20$, and a mass of $0.172\pm0.0004~M_{\odot}$. However, the final orbital period \citep[$0.44~\rm days$, see also Table 1 in][]{alth13} of the binary with a similar WD is below the measured value of PSR J1012+5307. The binary evolution did not take into account all works that can match the observed mass, radius, effective temperature, and surface gravity of the WD. The detailed binary evolution model given by \cite{mata20} can evolve to the present orbital period of PSR J1012+5307, however, their calculated effective temperature of the WD is higher than our results.

\begin{figure}
\centering
\includegraphics[width=1.15\linewidth,trim={0 0 0 0},clip]{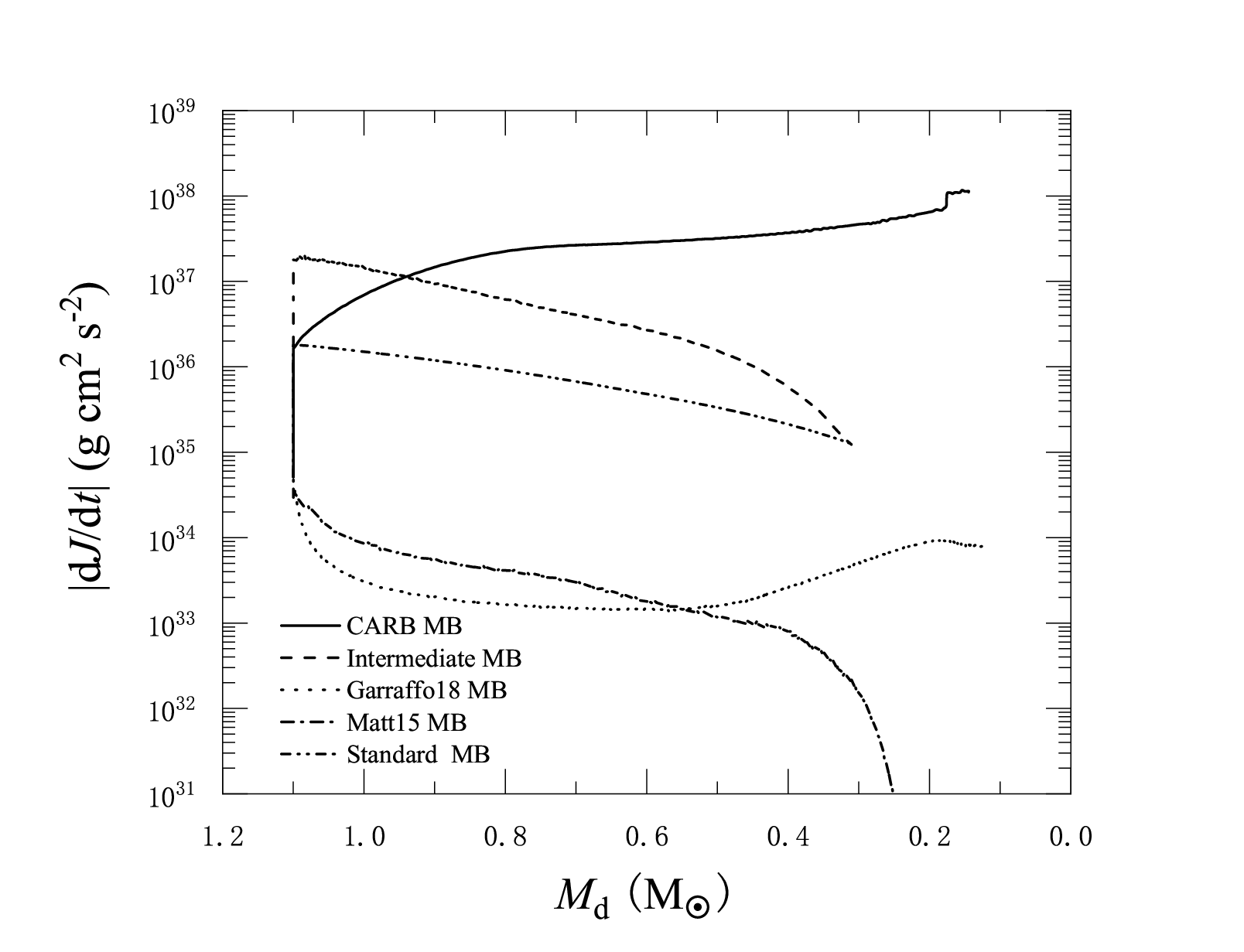}
\caption{Angular-momentum-loss rates as a function of the donor-star masses for different MB models. The initial parameters of our simulated NS binary are $M_{\rm NS,i}=1.50~M_{\rm \odot}$, $M_{\rm d,i}=1.10~M_{\rm \odot}$, and $P_{\rm orb,i}=1.1~{\rm days}$.} \label{fig:orbmass}
\end{figure}

\subsection{Discrepancy between the simulated and observed effective temperatures}
As in the previous section, our models cannot reproduce the effective temperature of the WD in PSR J1012+5307. In the CARB MB case, our simulated effective temperature is $T_{\rm eff}=11999~ \rm K$, which is higher than the observed value of $T_{\rm eff}=8362^{+25}_{-23}~ \rm K$. We alter the abundance of the donor star to be $Z=0.0142$, while the simulated $T_{\rm eff}$ is still high. In the H-R diagram, the donor star in our simulated NS binary has already evolved to the WD stage, in which the nascent WD is experiencing a rapid contraction and cooling process (see also Figure 3). To evolve to the present orbital period, our simulated cooling timescale of the WD is about $1.5-1.7~\rm Gyr$ for the CARB MB and 'Intermediate' MB models. According to $P=5.26~\rm ms$ and $\dot{P}=1.71\times10^{-20}~\rm s\,s^{-1}$ \citep{Alam21}, the spin-down age of the MSP in PSR J1012+5307 is $\tau_{\rm sd}=P/(2\dot{P})=4.9~\rm Gyr$. In principle, the spin-down age of the MSP should be consistent with the cooling timescale of the WD. This discrepancy may be caused by the contribution of H-shell flashes of the WD. These H-shell flashes in a close detached binary will lead to several temporary episodes of Roche lobe overflow in the ELM WD \citep{burd19,chen22}. A short-term mass transfer from the less massive ELM WD to the more massive NS can cause the orbit to temporarily widen, hence the NS-WD binary will spend a relatively long timescale to shrink to the present orbital period via gravitational radiation. As a consequence, the WD has a long timescale to cool to the observed effective temperature. On the other hand, the luminosity and effective temperature of the WD during the H-shell flashes will experience a wide range fluctuation, in which the effective temperature may match the observed value. However, both the luminosity and effective temperature change back to pre-flash values on a timescale of order 10 Myr. Such a short time window for observing PSR J1012+5307 in the H-shell flash stage with a low effective temperature is rather difficult \citep{chen22}. Therefore, a long cooling timescale caused by H-shell flashes of the WD may alleviate the discrepancy between the simulated and observed effective temperatures.


\section{Summary}
In this work, we explore the origin of the MSP PSR J1012+5307 with an ELM WD  to test different MB models. Employing a detailed stellar evolution model, we find that the five MB models can reproduce the observed orbital period and WD mass of PSR J1012+5307. However, the Matt15, Garraffo18, and standard MB models require a timescale longer than the Hubble time to evolve to the present orbital period and WD mass. For a binary system consisting of a $1.5~M_{\odot}$ NS and a MS donor star with a mass of $1.1~M_{\odot}$, the CARB MB and 'Intermediate' MB models can well reproduce the observed NS mass, WD mass, WD radius, WD surface gravity, and orbital period of PSR J1012+5307 when the initial orbital periods are $P_{\rm orb}=8.954$ and 25.2 days, respectively. However, our simulated effective temperature of the WD is higher than the observed value. This discrepancy may originate from our simulated short cooling timescale ($1.6-1.7~\rm Gyr$) of the WD, which is much shorter than the spin-down age ($4.9~\rm Gyr$) of the MSP of PSR J1012+5307. A long cooling timescale caused by H-shell flashes of the WD may alleviate the discrepancy between the simulated and observed effective temperatures.

\acknowledgments {We are extremely grateful to the anonymous referee for
helpful comments that improved this manuscript. This work was partly supported by the Major Science and Technology Program of Xinjiang Uygur Autonomous Region (No. 2022A03013-1), the National Key Research and Development Program of China (No. 2022YFC2205202), the National Natural Science Foundation of China (under grant Nos. 12273014, 12373044, and 12203051), and the Natural Science Foundation (under grant
No. ZR2021MA013) of Shandong Province.}

\end{document}